\documentclass{article}
\usepackage{graphicx} 
\usepackage{epsfig}
\usepackage[hyphens]{url}

\usepackage[margin=1in]{geometry}
\usepackage{fancyhdr}
\usepackage{paralist}
\usepackage{setspace}
\usepackage{verbatim}
\usepackage{authblk}

\usepackage[output-decimal-marker={.}]{siunitx}
\newcommand{\mxsecref}[1]{Section~\ref{sec:#1}}
\newcommand{\mxfigref}[1]{Fig.~\ref{fig:#1}}

\newcommand{\mxtabref}[1]{Table~\ref{tab:#1}}

\newcommand{\mxnewfig}{\begin{figure}[b]}
\newcommand{\mxendfig}{\end{figure}}

\newcommand{\mxcomponent}[3]{#1 & #2 & #3\\
\hline}

\newcommand{\mxurl}[1]{
\begin{center}
\url{#1}
\end{center}}

\title{Tools and Methodologies for System-Level Design}
\author[1]{Shuvra S.~Bhattacharyya}
\author[2]{Marilyn Wolf}
\affil[1]{Department of Electrical and Computer Engineering, and Institute for Advanced Computer Studies, University of Maryland at College Park, USA}
\affil[2]{School of Computing, University of Nebraska--Lincoln, USA}

\begin{document}

\maketitle

\section{Introduction}

System-level design, once the province of board designers, has now become a central concern for chip designers. Because chip design is a less forgiving design medium---design cycles are longer and mistakes are harder to correct---system-on-chip designers need a more extensive tool suite than may be used by board designers and a variety of tools and methodologies have been developed for system-level design of systems-on-chips (SoCs).

System-level design is less amenable to synthesis than are logic or physical design. As a result, system-level tools concentrate on {\em modeling, simulation, design space exploration, and design verification}. The goal of modeling is to correctly capture the system’s operational semantics, which helps with both implementation and verification.  The study of models of computation provides a framework for the description of digital systems. Not only do we need to understand a particular style of computation, such as dataflow, but we also need to understand how different models of computation can reliably communicate with each other. Design space exploration tools, such as hardware/software co-design, develop candidate designs to understand trade-offs. Simulation can be used not only to verify functional correctness but also to supply performance and power/energy information for design analysis. 

We will use two applications---video and neural networks---as examples in this chapter. Both are leading-edge applications that illustrate many important aspects of system-level design.  

The next two sections briefly introduce the applications and some system-on-chip architectures that may be the targets of system-level design tools. We will then study models of computation and languages for system-level modeling. We will then survey simulation techniques. We will close with discussions of hardware/software co-design and of machine learning for system design.

\section{Characteristics of Video Applications}

One of the primary uses of systems-on-chips for multimedia today is for video encoding---both compression and decompression. In this section we review the basic characteristics of video compression algorithms and the implications for video system-on-chip design. 

Video compression standards enable video devices to interoperate. The two major lines of video compression standards are MPEG and H.26x. The MPEG standards concentrate on broadcast applications, which allow for a more expensive compressor on the transmitter side in exchange for a simpler receiver. The H.26x standards were developed with videoconferencing in mind, in which both sides must encode and decode. The Advanced Video Codec (AVC) standard, also known as H.264, was formed by the confluence of the H.26x and MPEG efforts. H.264 is widely used in consumer video systems. 

Modern video compression systems combine lossy and lossless encoding methods to reduce the size of a video stream. Lossy methods throw away information such that the uncompressed video stream is not a perfect reconstruction of the original; lossless methods do allow the information provided to them to be perfectly reconstructed. Most modern standards use three major mechanisms:

\begin{itemize}
\item The discrete cosine transform (DCT) together with quantization. 
\item Motion estimation and compensation. 
\item Huffman-style encoding. 
\end{itemize}

Quantized DCT and motion estimation are lossy while Huffman encoding is lossless. These three methods leverage different aspects of the video stream’s characteristics to more efficiently encode it. 

The combination of DCT and quantization was originally developed for still images and is used in video to compress single frames. The discrete cosine transform is a frequency transform that turns a set of pixels into a set of coefficients for the spatial frequencies that form the components of the image represented by the pixels. The DCT is used over other transforms because a 2-D DCT can be computed using two 1-D DCTs, making it more efficient. In most standards the DCT is performed on an 8 x 8 block of pixels. The DCT does not itself lossily compress the image; rather, the quantization phase can more easily pick out information to throw away thanks to the structure of the DCT. Quantization throws out fine detail in the block of pixels, which correspond to the high-frequency coefficients in the DCT. The number of coefficients set to zero is determined on the level of compression desired. 

Motion estimation and compensation exploit the relationships between frames provided by moving objects. A reference frame is used to encode later frames through a motion vector, which describes the motion of a macroblock of pixels (16 x 16 in many standards). The block is copied from the reference frame into the new position described by the motion vector. The motion vector is much smaller than the block it represents. Two-dimensional correlation is used to determine the position of the macroblock’s position in the new frame; several positions in a search area are tested using 2-D correlation. An error signal encodes the difference between the predicted and the actual frames; the receiver uses that signal to improve the predicted picture. MPEG distinguishes several types of frames: I (inter) frames are not motion compensated; P (predicted) frames have been predicted from earlier frames; and B (bidirectional) frames have been predicted from both earlier and later frames. 

The results of these lossy compression phases are assembled into a bit stream and compressed using lossless compression such as Huffman encoding. This process reduces the size of the representation without further compromising image quality. 

It should be clear that video compression systems are actually heterogeneous collections of algorithms; this is also true of many other applications, including communications and security. A video computing platform must run several algorithms; those algorithms may perform very different types of operations, imposing very different requirements on the architecture. 

This has two implications for tools: first, that we need a wide variety of tools to support the design of these applications; second, the various models of computation and algorithmic styles used in different parts of an application must at some point be made to communicate to create the complete system. For example, DCT can be formulated as a dataflow problem thanks to its butterfly computational structure while Huffman encoding is often described in a control-oriented style. 

Several studies of multimedia performance on programmable processors have remarked on the significant number of branches in multimedia code. These observations contradict the popular notion of video as regular operations on streaming data. Fritts and Wolf~\cite{frit2000x1} measured the characteristics of the MediaBench benchmarks. 

They used path ratio to measure the percentage of instructions in a loop body that were actually executed. They found that the average path ratio of the MediaBench suite was 78\%, which indicates that a significant number of loops exercise data-dependent behavior. Talla et al.~\cite{tall2000x1} found that most of the available parallelism in multimedia benchmarks came from inter-iteration parallelism. Exploiting the complex parallelism found in modern multimedia algorithms requires that synthesis algorithms be able to handle more complex computations than simple ideal nested loops. 


The Google video coding unit (VCU) supports data center-scale video transcoding. An input video must be translated into several different formats:  multiple resolutions for display on devices ranging from high-end TVs to smartphones; multiple video formats to accommodate legacy devices~\cite{rang2021x1}. Multiple output transcoding (MOT) generates several output formats from a single input Group of Pictures (GOP). Transcoding is performed off-line, allowing non-causal video compression to be performed.  A first pipeline stage performs motion estimation, subblock partitioning, and rate distortion analysis for transform mode selection. A reference store in SRAM holds 768 x 512 pixels for motion estimation. Many of the memory accesses do not introduce hazards, the major exception occurring in in-loop deblocking filtering. A second stage performs entropy encoding. DRAM bandwidth was an important constraint in system design. The third stage performs loop filtering and frame buffer compression. 

\section{Platform Characteristics}

Many systems-on-chips are heterogeneous multiprocessors and the architectures designed for multimedia applications are no exception. In this section we review several systems-on-chips, including some general-purpose SoC architectures as well as several designed specifically for multimedia applications. 

Two very different types of hardware platforms have emerged for large-scale applications. On the one hand, many custom systems-on-chips have been designed for various applications. These custom SoCs are customized by loading software onto them for execution. On the other hand, platform FPGAs provide FPGA fabrics along with CPUs and other components; the design can be customized by programming the FPGA as well as the processor(s). These two styles of architecture represent different approaches for system-on-chip architecture and they require very different sorts of tools: custom SoCs require large-scale software support, while platform FPGAs are well-suited to hardware/software co-design. 

\subsection{Custom System-on-Chip Architectures}

The Texas Instruments OMAP processor (\url{http://www.omap.com}) is designed for mobile multimedia. It includes two processors, an ARM9 CPU and a TI C5x-series DSP, connected by a shared memory. OMAP implementations include a wide variety of peripherals. Some peripherals are shared between the two processors, including UARTs and mailboxes for interprocessor communication. Many peripherals, however, are owned by one processor. The DSP owns serial ports, DMA, timers, etc. The ARM9 owns the majority of the I/O devices, including USB, LCD controller, camera interface, etc. 

The Qualcomm Snapdragon 800~\cite{qual2014x1} is designed as a platform for smartphones. The chip includes a Krait 400 quad core CPU, graphics processing unit, DSP, 4G LTE modem, WiFi, and support for UltraHD video capture, playback and display.   

The Qorivva MPC5676R~\cite{free2012x1} is designed for automotive applications. It provides two Power Architecture cores, SIMD support for DSP and floating-point functions, a Time Processing Unit for high-rate periodic functions, and a FlexRay controller. 

\subsection{Graphics Processing Units (GPUs)}

Graphics processing units (GPUs) are widely used in desktop and laptop systems as well as smartphones.  GPUs are optimized for graphics rendering but have been applied to many other algorithms as well. GPUs provide SIMD-oriented architectures with floating-point support. 

\mxnewfig
\includegraphics[width=1.0\textwidth]
{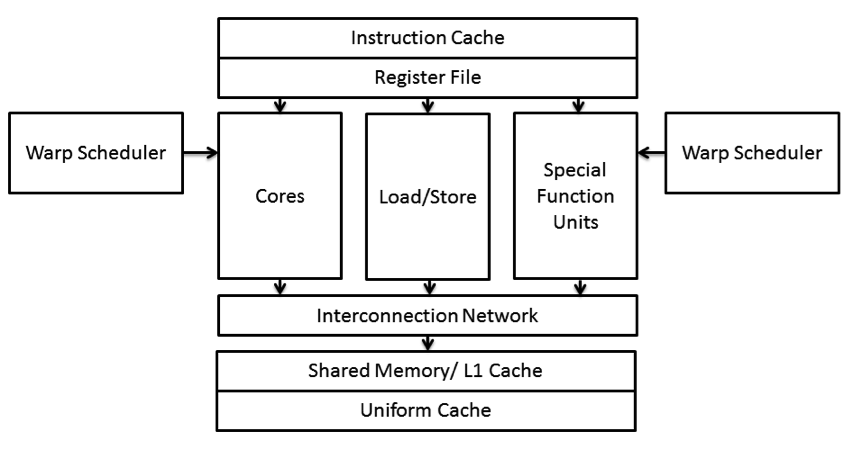}
\centering
 \caption{Organization of the Fermi GPU.}
 \label{fig:fermi}
\mxendfig

\mxfigref{fermi} shows the organization of the NVIDIA Fermi~\cite{nvid2009x1}. Three types of processing elements are provided: cores, each of which has a floating-point and an integer unit; load/store units; and special function units. A hierarchy of register files, caches, and shared memory provide very high memory bandwidth. A pair of warp processors controls operation. Each warp scheduler can control a set of 32 parallel threads.

\subsection{Platform FPGAs}

Field-programmable gate arrays (FPGAs)~\cite{wolf2004x1} have been used for many years to implement logic designs. The FPGA provides a more general structure than programmable logic devices, allowing denser designs. They are less energy-efficient than custom ASICs but do not require the long ASIC design cycle. 

Many FPGA design environments provide small, customizable {\em soft processors} that can be embedded as part of the logic in an FPGA. Examples include the Xilinx MicroBlaze and Altera Nios II. Nios II supports memory management and protection units, separate instruction and data caches, pipelined execution with dynamic branch prediction, up to 256 custom instructions, and hardware accelerators. MicroBlaze supports memory management units, instruction and data caches, pipelined operation, floating-point operations, and instruction set extensions. 

The term {\em programmable SoC} refers to an FPGA that provides one or more hard logic CPUs in addition to a programmable FPGA fabric. Platform FPGAs provide a very different sort of heterogeneous platform than custom SoCs. The FPGA fabric allows the system designer to implement new hardware functions. While they generally do not allow the CPU itself to be modified, the FPGA logic is a closely coupled device with high throughput to the CPU and to memory. The CPU can also be programmed using standard tools to provide functions that are not well suited to FPGA implementation. The Xilinx Zynq 7000~\cite{xili2014x1} incorporates dual ARM Cortex-A9 with Neon engines as well as programmable logic and a variety of I/O devices. The Altera Arria 10 incorporates a dual-core ARM Cortex-A9 MPCore, programmable logic, and a variety of I/O devices~\cite{alte2014x1}. The Cypress PSoC 5LP~\cite{cypr2013x1} provides an ARM Cortex-M3 processor, universal digital blocks, and both analog and digital peripherals. Mixed-signal peripherals can be configured on the PSoC and connected via interrupts to the CPU.

\subsection{Neural Network Accelerators}


The Google Tensor Processing Unit (TPU) was designed to execute neural network inference.  Neural networks perform a combination of matrix operations and non-linear function evaluations~\cite{joup2017x1}.  The matrix operations take the form of tensors due to the existence of multiple sets of coefficients. Unlike the motion estimation algorithms of video coding, which typically operate on 8 x 8 matrices, neural networks support a wide range of tensor sizes.  The TPU is designed to efficiently operate on a range of tensor sizes found in practical neural network architectures. The matrix multiply unit contains 256 x 256 multiply-accumulate units that feed an array of 32-bit accumulators. Each accumulator holds a 256-element partial sum. The accumulator block contains 4096 accumulators. A FIFO feeds weights to the matrix unit. The TPU is controlled by CISC instructions, typically 10-20 clock cycles long. Instructions perform operations such as reading and writing host memory, reading weights, matrix multiplication and convolution, and activation functions The TPU is attached to a host processor on a PCIe bus; the host sends instructions to the TPU for execution.  

\section{Abstract Design Methodologies}

Several groups have developed abstract models for system-level design methodologies. These models help us to compare concrete design methodologies. 

\mxnewfig
\includegraphics[width=1.0\textwidth]
{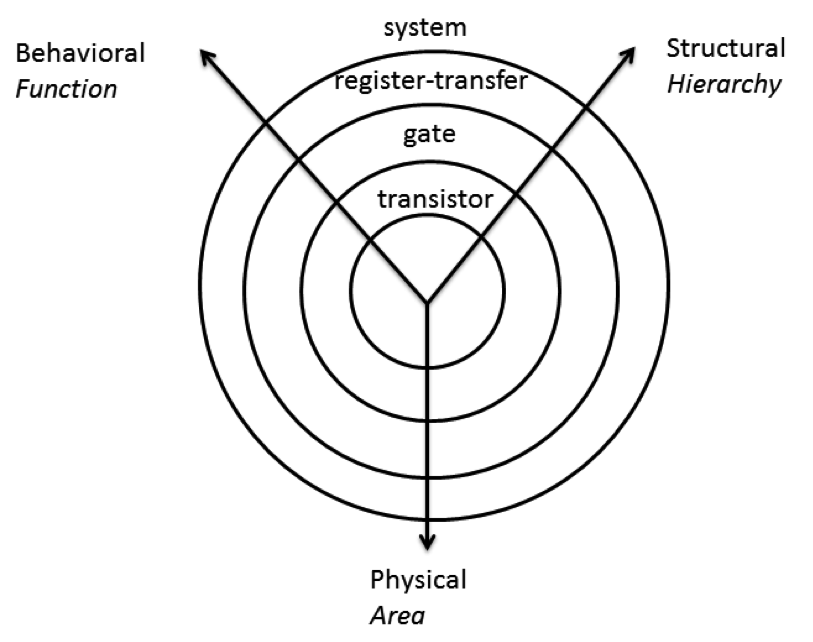}
\centering
 \caption{ The Y-chart model for design methodologies~\cite{gajs1983x1}.}
 \label{fig:ychart}
\mxendfig

An early influential model for design methodologies was the Gajski-Kuhn Y-Chart~\cite{gajs1983x1}. As shown in \mxfigref{ychart}, the model covers three types of refinement (structural, behavioral, and physical) at four levels of design abstraction (transistor, gate, register-transfer, system). A design methodology could be viewed as a trajectory through the Y-chart as various refinements are performed at different levels of abstraction. 


\mxnewfig
\includegraphics[width=1.0\textwidth]
{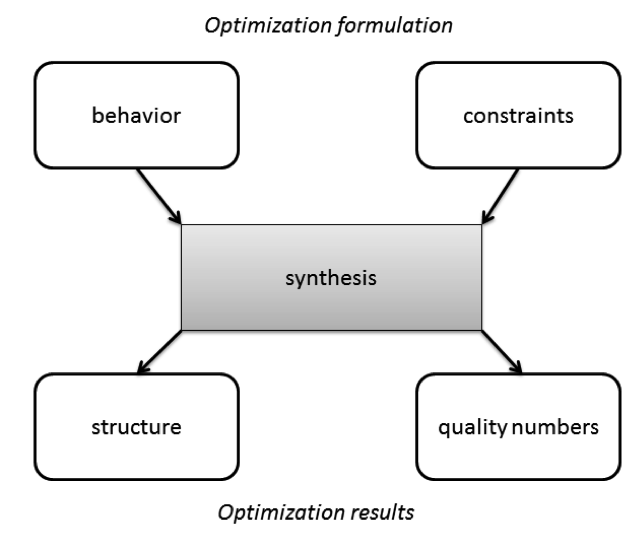}
\centering
 \caption{The X-Chart model for design methodologies~\cite{gers2009x1}.}
 \label{fig:xchart}
\mxendfig

The X-Chart model~\cite{gers2009x1} has been proposed as a model for system-on-chip design methodologies. As shown in \mxfigref{xchart}, a system specification is given by the combination of a behavioral description that describe the system function and a set of constraints that describes the non-functional requirements on the design. A synthesis procedure generates a structural implementation and a set of quality numbers by which the structure can be judged. 


\section{Model-based Design Methodologies}

Increasingly, developers of hardware and software for embedded computer systems are viewing aspects of the design and implementation processes in terms of domain-specific models of computation. Models of computation provide formal principles that govern how functional components in a computational specification operate and interact (e.g., see~\cite{lee1998x3}). A domain-specific model of computation is designed to represent applications in a particular functional domain such as digital signal, image and video processing (DSP); control system design; machine learning; communication protocols or more general classes of discrete, control-flow intensive decision-making processes; graphics; and device drivers. For discussions of some representative languages and tools that are specialized for these application domains, see for example~\cite{basu1997x1, kons1994x1, lauw1995x1, lee1989x2, conw2004x1, mani1998x2, prou2001x1, thib1999x1, bhat2019x1, eker2012x1, pelc2013x1, bhat2008x4, edwa2000x1}, and some more recent ones that are covered in \mxsecref{languages}.

Processors expose a low-level Turing model at the instruction set. Traditional “high-level” programming languages like C, C++, and Java maintain the essential elements of that Turing model, including imperative semantics and memory-oriented operation. Mapping the semantics of modern, complex applications onto these low-level models is both time-consuming and error-prone.  As a result, new programming languages and their associated design methodologies have been developed to support applications such as signal/image processing and communications. Compilers for these languages provide correct-by-construct translation of application-level operations to the Turing model, which both improves designer productivity and provides a stronger, more tool-oriented verification path~\cite{jant2003x1}. 

\subsection{Dataflow Models}
\label{sec:dataflow}

For most signal processing and machine learning (SPML) applications, a significant part of the computational structure is well-suited to modeling in a dataflow model of computation. In the context of programming models, dataflow refers to a modeling methodology where computations are represented as directed graphs in which vertices (actors) represent functional components and edges between actors represent first-in-first-out (FIFO) channels that buffer data values (tokens) as they pass from an output of one actor to an input of another. Dataflow actors can represent computations of arbitrary complexity; typically in DSP design environments they are specified using conventional languages such as C or assembly language, and their associated tasks range from simple, ``fine-grained'' functions such as addition and multiplication to “coarse-grain” DSP kernels or subsystems such as FFT units and adaptive filters.  

The development of application modeling and analysis techniques based on dataflow graphs was inspired by the computation graphs of Karp and Miller~\cite{karp1966x1}, and the process networks of Kahn~\cite{kahn1974x1}. A unified formulation of dataflow modeling principles as they apply to DSP design environments is provided by the dataflow process networks model of computation of Lee and Parks~\cite{lee1995x1}. 

A dataflow actor is enabled for execution any time it has sufficient data on its incoming edges (i.e., in the associated FIFO channels) to perform its specified computation. An actor can execute at any time when it is enabled (data-driven execution). In general, the execution of an actor results in some number of tokens being removed (consumed) from each incoming edge, and some number being placed (produced) on each outgoing edge. This production activity in general leads to the enabling of other actors. 

The order in which actors execute, called the {\em schedule}, is not part of a dataflow specification, and is constrained only by the simple principle of data-driven execution defined above. This is in contrast to many alternative computational models, such those that underlie procedural languages, in which execution order is ``over-specified'' by the programmer~\cite{ambl1992x1}. The schedule for a dataflow specification may be determined at compile time (if sufficient static information is available), at run-time, or using a mixture of compile-time and run-time techniques. A particularly powerful class of scheduling techniques, referred to as {\em quasi-static scheduling} (e.g., see~\cite{ha1991x1}), involves most, but not all of the scheduling decisions being made at compile time.

\mxfigref{videoDataflow} shows an illustration of a video processing subsystem that is modeled using dataflow semantics. This is a design, developed using the Ptolemy II tool for model-based embedded system design~\cite{eker2003x4}, of an MPEG-2 subsystem for encoding the P frames that are processed by an enclosing MPEG-2 encoder system. A thorough discussion of this MPEG-2 system and its comparison to a variety of other modeling representations is presented in~\cite{ko2005x1}. The components in the design of \mxfigref{videoDataflow} include actors for the discrete cosine transform, zig-zag scanning, quantization, motion compensation, and run length coding. The arrows in the illustration correspond to the edges in the underlying dataflow graph.  

The actors and their interactions all conform to the semantics of {\em synchronous dataflow} ({\em SDF}), which is a restricted form of dataflow that is efficient for describing a broad class of DSP applications and has particularly strong formal properties and optimization advantages~\cite{lee1987x1, bhat2019x1}. Specifically, SDF imposes the restriction that the number of tokens produced and consumed by each actor on each incident edge is constant. Many commercial DSP design tools have been developed that employ semantics that are equivalent to or closely related to SDF. Examples of such tools include Agilent’s SystemVue, Kalray’s MPPA Software Development Kit, National Instrument’s LabVIEW, and Synopsys’s SPW. Simulink, another widely used commercial tool, also exhibits some relationships to the SDF model (e.g., see~\cite{lubl2008x2}). 

\subsection{Dataflow Modeling for Video Processing}

\mxnewfig
\includegraphics[width=1.0\textwidth]
{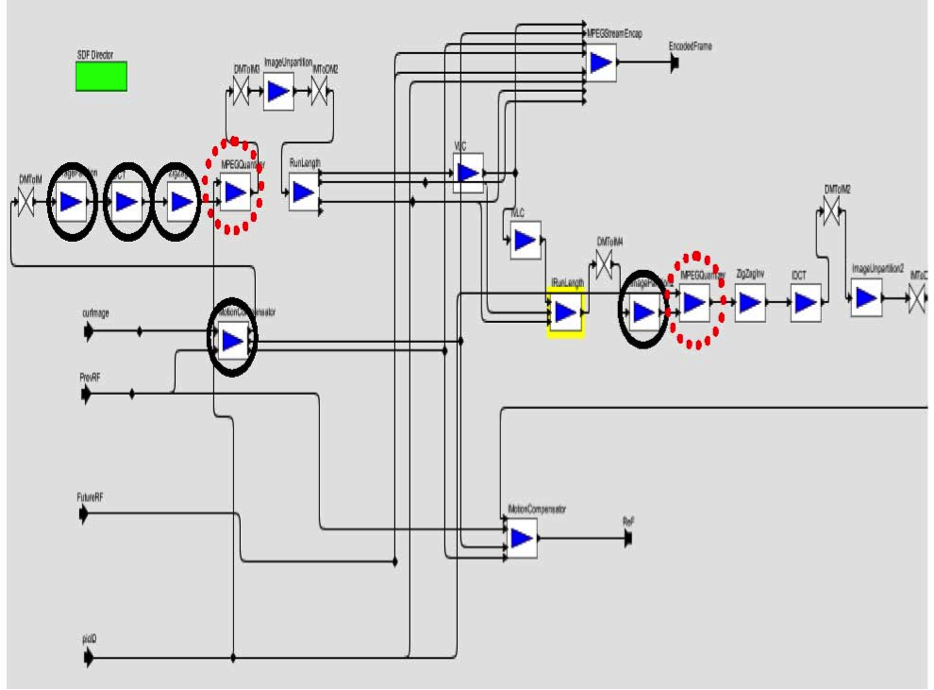}
\centering
 \caption{A video processing system modeled in dataflow.}
 \label{fig:videoDataflow}
\mxendfig

In the context of video processing, SDF permits accurate representation of many useful subsystems, such as the P-frame encoder shown in \mxfigref{videoDataflow}. However, such modeling is often restricted to a highly coarse level of granularity, where actors process individual frames or groups of successive frames on each execution. Modeling at such a coarse granularity can provide compact, top-level design representations, but greatly limits the benefits offered by the dataflow representation since most of the computation is subsumed by the general-purpose, intra-actor program representation. For example, the degree of parallel processing and memory management optimizations exposed to a dataflow-based synthesis tool becomes highly limited at such coarse levels of actor granularity. An example of a coarse grain dataflow actor that “hides” significant amounts of parallelism is the DCT actor in \mxfigref{videoDataflow}. 

\subsection{Multidimensional Dataflow Models}

A number of alternative dataflow modeling methods have been introduced to address this limitation of SDF modeling for video processing and more generally, multi-dimensional signal processing applications. For example, the multi-dimensional synchronous dataflow (MD-SDF) model extends synchronous dataflow semantics to allow constant-sized, $n$-dimensional vectors of data to be transferred across graph edges, and provides support for arbitrary sampling lattices and lattice-changing operations~\cite{murt2002x1}. Intuitively, a sampling lattice can be viewed as a generalization to multiple dimensions of a uniformly spaced configuration of sampling points for a one-dimensional signal~\cite{vaid1993x1}; hexagonal and rectangular lattices are examples of commonly-used sampling lattices for two-dimensional signals. The computer vision synchronous dataflow (CV-SDF) model is designed specifically for computer vision applications, and provides a notion of {\em structured buffers} for decomposing video frames along graph edges; accessing neighborhoods of image data from within actors, in addition to the conventional production and consumption semantics of dataflow; and allowing actors to efficiently access previous frames of image data~\cite{stic2002x2, kein2006x1}. Blocked dataflow (BLDF) is a meta-modeling technique for efficiently incorporating hierarchical, block-based processing of multi-dimensional data into a variety of dataflow modeling styles, including SDF and MD-SDF~\cite{ko2005x1}. Windowed synchronous dataflow (WSDF) is a model of computation that deeply integrates support for sliding window algorithms into the framework of static dataflow modeling~\cite{kein2006x1}. Such support is important in the processing of images and video streams, where sliding window operations play a fundamental role. 

\subsection{Control Flow}

As described previously, modern video processing applications are characterized by some degree of control flow processing for carrying out data-dependent configuration of application tasks and changes across multiple application modes. For example, in MPEG-2 video encoding, significantly different processing is required for I frames, P frames, and B frames. Although the processing for each particular type of frame (I, P, or B) conforms to the SDF model, as illustrated for P frame processing in \mxfigref{videoDataflow}, a layer of control flow processing is needed to efficiently integrate these three types of processing methods into a complete MPEG-2 encoder design. The SDF model is not well-suited for performing this type of control flow processing and more generally for any functionality that requires dynamic communication patterns or activation/deactivation across actors.  

A variety of alternative models of computation have been developed to address this limitation, and integrate flexible control flow capability with the advantages of dataflow modeling. In Buck’s Boolean dataflow model~\cite{buck1993x1} and the subsequent generalization as integer-controlled dataflow~\cite{buck1994x2}, provisions for such flexible processing were incorporated without departing from the framework of dataflow, and in a manner that facilitates construction of efficient quasi-static schedules. In Boolean dataflow, the number of tokens produced or consumed on an edge is either fixed or is a two-valued function of a control token present on a control terminal of the same actor. It is possible to extend important SDF analysis techniques to Boolean dataflow graphs by employing symbolic variables. In particular, in constructing a schedule for Boolean dataflow actors, Buck’s techniques attempt to derive a quasi-static schedule, where each conditional actor execution is annotated with the run-time condition under which the execution should occur. Boolean dataflow is a powerful modeling technique that can express arbitrary control flow structures; however, as a result, key formal verification properties of SDF, such as bounded memory and deadlock detection, are lost (the associated analysis problems are not decidable) in the context of general Boolean dataflow specifications. 

\subsection{Integration with Finite-State Machine and Mode-based Modeling Methods}
\label{sec:FSMintegration}

In recent years, several modeling techniques have also been proposed that enhance expressive power by providing precise semantics for integrating dataflow or dataflow-like representations with finite state machine (FSM) models and related methods for specifying and transitioning between different modes of actor behavior. These include El Greco~\cite{buck2000x1}, which evolved into the Synopsys System Studio, and provides facilities for ``control models'' to dynamically configure specification parameters; *charts (pronounced “starcharts”) with heterochronous dataflow as the concurrency model~\cite{gira1999x1}; the FunState intermediate representation~\cite{thie1999x1}; the DF* framework developed at K. U. Leuven~\cite{coss2000x1}; the control flow provisions in bounded dynamic dataflow~\cite{pank1994x1}; enable-invoke dataflow~\cite{plis2008x5}; and scenario-aware dataflow (SADF)~\cite{thee2006x1}. 

\subsection{Video Processing Examples}
\label{sec:videoExamples}

\mxfigref{mpeg2} shows an illustration of a model of a complete MPEG-2 video encoder system that is constructed using Ptolemy, builds on the P-frame-processing subsystem of \mxfigref{videoDataflow}, and employs multiple dataflow graphs nested within a finite state machine representation. Details on this application model can be found in~\cite{ko2005x1}.  

\mxnewfig
\includegraphics[width=1.0\textwidth]
{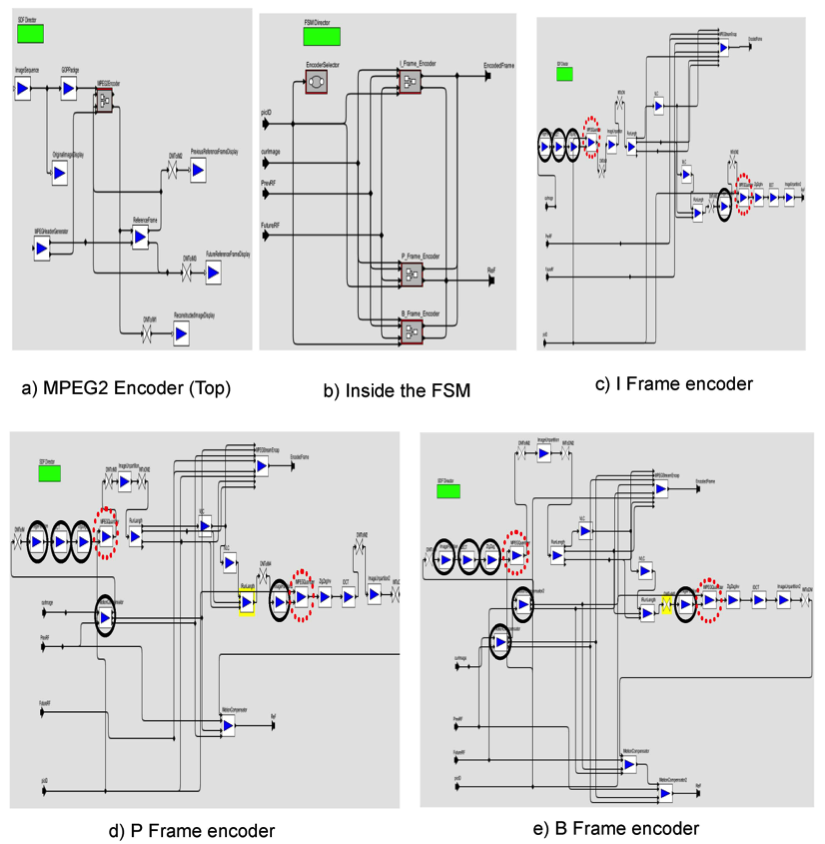}
\centering
 \caption{An MPEG-2 video encoder specification.}
 \label{fig:mpeg2}
\mxendfig

\mxfigref{mpeg4} shows a block diagram, adapted from~\cite{thee2006x1}, of an MPEG-4 decoder that is specified in terms of SADF. Descriptive names for the actors in this example are listed in \mxtabref{mpegComponents}, along with their SADF component types, which are either shown as ``K'' for kernel or ``D'' for detector. Intuitively, kernels correspond to data processing components of the enclosing dataflow graph, while detectors are used for control among different modes of operation (``scenarios'') for the kernels.  The FD actor in the example of \mxfigref{mpeg4} determines the frame type (I or P frame), and is designed as a detector. The other actors in the specification are kernels. For more details on this example and the underlying scenario-aware dataflow model of computation, we refer the reader to~\cite{thee2006x1}. The``D'' symbols that appear next to some of the edges in \mxfigref{mpeg4} represent dataflow {\em delays}, which correspond to initial tokens on the edges. 

\mxnewfig
\includegraphics[width=1.0\textwidth]
{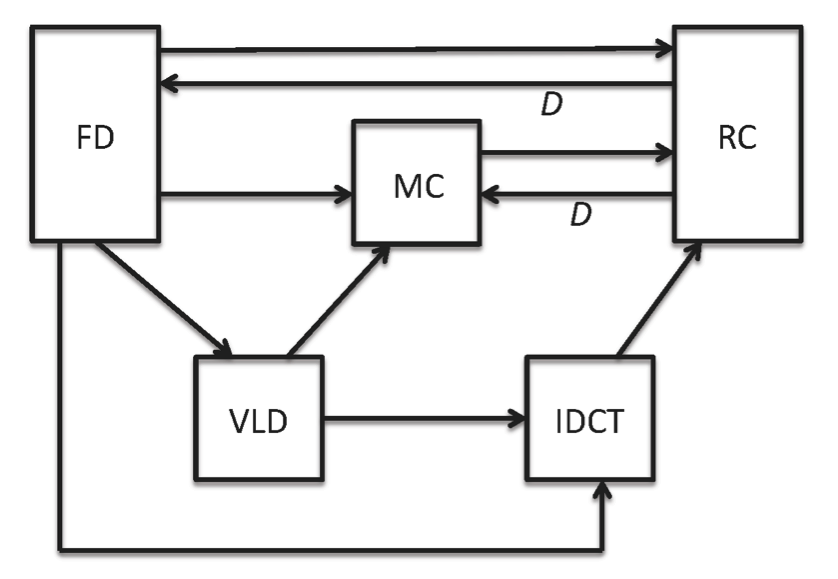}
\centering
 \caption{A block diagram, adapted from~\cite{thee2006x1}, of an MPEG-4 decoder that is specified in terms of 
  scenario-aware dataflow.}
 \label{fig:mpeg4}
\mxendfig

\begin{table}[b]
  \centering
  \begin{tabular}{|c|l|l|}
  \hline
  \mxcomponent{\bf Abbreviation}{\bf Descriptive Name}{\bf Type}
  \mxcomponent{FD}{Frame Detector}{D}
  \mxcomponent{IDCT}{Inverse Discrete Cosine Transformation}{K}
  \mxcomponent{MC}{Motion Compensation}{K}
  \mxcomponent{RC}{Reconstruction}{K}
  \mxcomponent{VLD}{Variable Length Decoding}{K}
  \end{tabular}
  \caption{Modeling components used in the MPEG-4 decoder example of \mxfigref{mpeg4}.}
  \label{tab:mpegComponents}
\end{table}

\section{Languages and Tools for Model-based Design}
\label{sec:languages}

In this section we survey research on tools for model-based design of embedded systems, with an emphasis on tool capabilities that are relevant to video processing systems and machine learning systems. We discuss several representative tools that employ established and experimental models of computation, and provide features for simulation, rapid prototyping, synthesis, and optimization. For more extensive coverage of model-based design tools for video processing systems and related application areas, we refer the reader to~\cite{bhat2019x1}.

\subsection{CAL}

CAL is a language for dataflow programming that has been applied to hardware and software synthesis and a wide variety of applications, with a particular emphasis on applications in video processing~\cite{eker2012x1}. One of the most important applications of CAL to date is the incorporation of a subset of CAL, called RVC-CAL, as part of the MPEG reconfigurable video coding (RVC) standard~\cite{jann2010x1}. In CAL, dataflow actors are specified in terms of entities that include {\em actions}, {\em guards}, {\em port patterns}, and {\em priorities}. An actor can contain any number of actions, where each action describes a specific computation that is to be performed by the actor, including the associated consumption and production of tokens at the actor ports, when the action is executed. Whether or not an action can be executed at any given time depends in general on the number of available input tokens, the token values, and the actor state. These fireability conditions are specified by input patterns and guards of the action definition. The relatively high flexibility allowed for constructing firing conditions makes CAL a very general model, where fundamental scheduling-related problems become undecidable, as with Boolean dataflow and other highly expressive, ``dynamic dataflow'' models.

Input patterns also declare variables that correspond to input tokens that are consumed when the action executes, and that can be referenced in specifying the computation to be performed by the action. Such deep integration of dataflow-based, actor interface specification with specification of the detailed internal computations performed by an actor is one of the novel aspects of CAL.   

Priorities in CAL actor specifications provide a way to select subsets of actions to execute when there are multiple actions that match the fireability conditions defined by the input patterns and guards.  For more details on the CAL language, we refer the reader to the CAL language report~\cite{eker2003x2}. 

A wide variety of tools has been developed to support design of hardware and software systems using CAL. For example, OpenDF was introduced in~\cite{bhat2008x4} as an open-source simulation and compilation framework for CAL; the open RVC-CAL compiler (Orcc) is an open source compiler infrastructure that enables code generation for a variety of target languages and platforms~\cite{yviq2013x2}; Boutellier et al. present a plug-in to OpenDF for multiprocessor scheduling of RVC systems that are constructed using CAL actors~\cite{bout2008x1}; and Gu et al. present a tool that automatically extracts and exploits statically-schedulable regions from CAL specifications~\cite{gu2011x2}. 

\subsection{Compaan}

MATLAB is one of the most popular programming languages for algorithm development, and high-level functional simulation for DSP applications. In the Compaan project, developed at Leiden University, systematic techniques have been developed for synthesizing embedded software and FPGA-based hardware implementations from a restricted class of MATLAB programs known as parameterized, static nested loop programs~\cite{stef2004x1}. In Compaan, an input MATLAB specification is first translated into an intermediate representation based on the Kahn process network model of computation~\cite{kahn1974x1}. The Kahn process network model is a general model of data-driven computation that subsumes as a special case the dataflow process networks mentioned earlier in this chapter. Like dataflow process networks, Kahn process networks consist of concurrent functional modules that are connected by FIFO buffers with non-blocking writes and blocking reads; however, unlike the dataflow process network model, modules in Kahn process networks do not necessarily have their execution decomposed a priori into well-defined, discrete units of execution~\cite{lee1995x1}. 

Through its aggressive dependence analysis capabilities, Compaan combines the widespread appeal of MATLAB at the algorithm development level with the guaranteed determinacy, compact representation, simple synchronization, and distributed control features of Kahn process networks for efficient hardware/software implementation. 

Technically, the Kahn process networks derived in Compaan can be described as equivalent cyclo-static dataflow graphs~\cite{bils1996x1, depr2006x1}, and therefore fall under the category of dataflow process networks. However, these equivalent cyclo-static dataflow graphs can be very large and unwieldy to work with, and therefore, analysis in terms of the Kahn process network model is often more efficient and intuitive. 

Development of the capability for translation from MATLAB to Kahn process networks was originally developed by Kienhuis, Rijpkema, and Deprettere~\cite{kien2000x1}, and this capability has since evolved into an elaborate suite of tools for mapping Kahn process networks into optimized implementations on heterogeneous hardware/software platforms consisting of embedded processors and FPGAs~\cite{stef2004x1}. Among the most interesting optimizations in the Compaan tool suite are dependence analysis mechanisms that determine the most specialized form of buffer implementation, with respect to reordering and multiplicity of buffered values, for implementing inter-process communication in Kahn process networks~\cite{turj2004x1}. 

At Leiden University, Compaan was succeeded by the Daedalus project, which provides a design flow for mapping embedded multimedia applications onto multiprocessor system-on-chip devices~\cite{niko2008x2}.  

\subsection{PREESM}

PREESM (Parallel and Real-time Embedded Executives Scheduling Method) is an extensible, Eclipse-based framework for rapid programming of signal processing systems~\cite{pelc2013x1, pelc2009x2}. Special emphasis is placed in PREESM for integrating and experimenting with different kinds of multiprocessor scheduling techniques and associated target architecture models. Such modeling and experimentation is useful in the design and implementation of real-time video processing systems, which must often satisfy stringent constraints on latency, throughput, and buffer memory requirements. 

Various types of tools for compilation, analysis, scheduling and architecture modeling can be integrated into PREESM as Eclipse~\cite{holz2004x1} plug-ins. Existing capabilities of PREESM emphasize use of architecture models and scheduling techniques that are targeted to mapping applications onto Texas Instruments programmable digital signal processors, including the TMS320C64x+ series of processors. Applications are modeled in PREESM using SDF graphs, while target architectures are modeled as interconnections of abstracted processor cores, hardware co-processors and communication media. Both homogeneous and heterogeneous architectures can be modeled, and emphasis also is placed on careful modeling of DMA-based operation associated with the communication media. 

 Multiprocessor scheduling of actors in PREESM is performed using a form of list scheduling. A randomized version of the list scheduling algorithm is provided based on probabilistic generation of refinements to the schedule derived by the basic list scheduling technique. This randomized version can be executed for an arbitrary amount of time, as determined by the designer, after which the best solution observed during the entire execution is returned. Additionally, the randomized scheduling algorithm can be used to initialize the population of a genetic algorithm, which provides a third alternative for multiprocessor scheduling in PREESM. A plug-in for “edge scheduling” is provided within the scheduling framework of PREESM to enable application of alternative methods for mapping interprocessor communication operations across the targeted interconnection of communication media. 

 Pelcat et al. present a study involving the application of PREESM to rapid prototyping of a stereo vision system~\cite{pelc2014x2}. 

 \subsection{Ptolemy}

 The Ptolemy project at U.~C.~Berkeley has had considerable influence on the general trend toward viewing embedded systems design in terms of models of computation~\cite{eker2003x4, buck1994x1}. The design of Ptolemy emphasizes efficient modeling and simulation of embedded systems based on the interaction of heterogeneous models of computation. A key motivation is to allow designers to represent each subsystem of a design in the most natural model of computation associated with that subsystem, and allow subsystems expressed in different models of computation to be integrated seamlessly into an overall system design.  

 A key constraint imposed by the Ptolemy approach to heterogeneous modeling is the concept of {\em hierarchical heterogeneity}~\cite{buck1994x1}. It is widely understood that in hierarchical modeling, a system specification is decomposed into a set $C$ of subsystems in which each subsystem can contain one or more hierarchical components, each of which represents another subsystem in $C$. Under hierarchical heterogeneity, each subsystem in $C$ must be described using a uniform model of computation, but the nested subsystem associated with a hierarchical component $H$ can be expressed (refined) in a model of computation that is different from the model of computation that expresses the subsystem containing $H$.

 Thus, under hierarchical heterogeneity, the integration of different models of computation must be achieved entirely through the hierarchical embedding of heterogeneous models. A key consequence is that whenever a subsystem $S_1$ is embedded in a subsystem $S_2$ that is expressed in a different model of computation, the subsystem $S_1$ must be abstracted by a hierarchical component in $S_2$ that conforms to the model of computation associated with $S_2$. This provides precise constraints for interfacing different models of computation. Although these constraints may not always be easy to conform to, they provide a general and unambiguous convention for heterogeneous integration and perhaps even more importantly, the associated interfacing methodology allows each subsystem to be analyzed using the techniques and tools available for the associated model of computation. 

Ptolemy has been developed through a highly flexible, extensible, and robust software design, and this has facilitated experimentation with the underlying modeling capabilities by many research groups in various aspects of embedded systems design.  Major areas of contribution associated with development of Ptolemy that are especially relevant for video processing systems include hardware/software codesign, as well as contributions in dataflow-based modeling, analysis, and synthesis (e.g., see~\cite{murt2002x1, kala1993x1, neue2004x2, zhou2007x1}).  

The original version of Ptolemy was succeeded by a new incarnation, called Ptolemy II, which is a Java-based tool that furthers the application of model-based design and hierarchical heterogeneity~\cite{eker2003x4}, and provides an even more malleable software infrastructure for experimentation with new techniques involving models of computation.  

An important theme in Ptolemy II is the reuse of actors across multiple computational models. Through an emphasis in Ptolemy II on support for {\em domain polymorphism}, the same actor definition can in general be applicable across a variety of models of computation. In practice, domain polymorphism greatly increases reuse of actor code. Techniques based on interface automata~\cite{alfa2001x1} have been developed to systematically characterize the interactions between actors and models of computation, and reason about their compatibility (i.e., whether or not it makes sense to instantiate an actor in specifications that are based on a given model)~\cite{lee2001x4}. 

\subsection{SysteMoc}

SysteMoc is a SystemC-based library for dataflow-based, hardware/software co-design and synthesis of signal processing systems. (See \mxsecref{simulation} for more details about SystemC, the simulation language on which SysteMoc was developed.) SysteMoc is based on a form of dataflow in which the model for each actor $A$ includes a set $F$ of functions, and a finite state machine (FSM), called the firing FSM of $A$. Each function $f \in F$ is classified as either an {\em action} function or a {\em guard} function. The action functions provide the core data processing capability of the actor, while the guard functions determine the activation of transitions in the firing FSM. Guard functions can access values of tokens present at the input edges of an actor (without consuming them), thereby enabling data-dependent sequencing of actions through the firing FSM. Furthermore, each transition $t$ of the firing FSM has an associated action function $x(t) \in F$, which is executed when the transition $t$ is activated. Thus, SysteMoc provides an integrated method for specifying, analyzing and synthesizing actors in terms of FSMs that control sets of alternative dataflow behaviors. 

SysteMoc has been demonstrated using a design space exploration case study for FPGA-based implementation of a two-dimensional inverse discrete cosine transform, as part of a larger case study involving an MPEG-4 decoder~\cite{haub2007x1}.This case study considered a five-dimensional design evaluation space encompassing throughput, latency, number of look-up tables (LUTs), number of flip-flops, and a composite resource utilization metric involving the sum of block RAM (BRAM) and multiplier resources.  Among the most impressive results of the case study was the accuracy with which the design space exploration framework developed for SysteMoc was able to estimate FPGA hardware resources. For more details on SysteMoc and the MPEG-4 case study using SysteMoc, we refer the reader to~\cite{haub2007x1}. 

\subsection{fpgaConvNet}

The fpgaConvNet framework provides design automation for mapping of convolutional neural networks (CNNs) onto FPGAs~\cite{veni2019x1}. The framework handles different varieties of CNNs, including dense, inception-based, and residual networks. A distinguishing characteristic of the framework is its unified formulation of the design space in terms of the SDF model of computation (see \mxsecref{dataflow}) and linear algebra concepts, and significant extensions to SDF modeling that it incorporates to capture details of digital hardware mapping.  The framework takes as input a trained CNN model specified using Torch or Caffe, and a specification of the amount of each type of available FPGA resource, including the numbers of DSP blocks, lookup tables, flip flops and block RAMs, and the off-chip memory capacity. The off-chip memory bandwidth is also included in the platform model.

A hardware architecture (design point) for implementing a CNN is represented as an SDF graph, where each layer in the CNN is represented as a linear connection of SDF actors. Each actor is taken from a library of pre-defined building blocks that are used to construct CNN layers. For example, \mxfigref{cnnLayer} shows the SDF subgraph associated with a convolutional layer in terms of five building blocks. Here, each actor $A, B, C, D, E$ respectively represents a building block that reads from memory, produces sliding windows across the data read from memory, copies data to parallel streams to enable data-parallel operation, applies a bank of parallel convolution units, and writes output data to memory. The production and consumption rates along the edges are represented in terms of the memory bandwidth $Z$, number of convolutional units $N$, and the filter width/height $K$.

\mxnewfig
\includegraphics[width=1.0\textwidth]
{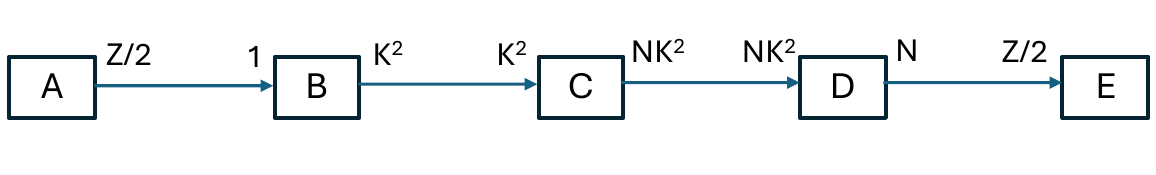}
\centering
 \caption{An an example of an SDF subgraph corresponding to a CNN layer in fpgaConvNet.}
 \label{fig:cnnLayer}
\mxendfig

To expose hardware parameters in the design space formulation, the SDF topology matrix $\Gamma$ is decomposed into the Hadamard product (element-wise product) of three matrices $\Gamma = \Gamma_s \odot \Gamma_c \odot \Gamma_r$ --- the streams matrix, channels matrix, and rate matrix. The SDF topology matrix $\Gamma$ is a fundamental representation for analysis of SDF graphs~\cite{lee1987x1}. Given an SDF graph $G$, columns of the associated topology matrix $\Gamma$ are indexed by the vertices (actors) of $G = (V, E)$ and the rows are indexed by the edges. For each edge $e = (x, y)$, $\Gamma[e, x]$ is equal to the production rate of actor $x$ onto $e$, and $\Gamma[e, y]$ is equal to the negative of the consumption rate of $y$ from $e$. For all $v \in V$ and $e \in E$ such that there is no connection between $v$ and $e$, $\Gamma[v, x] = 0$.

In the decomposed form of the topology matrix $\Gamma = \Gamma_s \odot \Gamma_c \odot \Gamma_r$, the streams matrix $\Gamma_s$ gives the number of parallel streams associated with each edge. The channels matrix $\Gamma_c$ gives the width of each stream in words along with the direction of dataflow, where positive and negative values correspond to different directions of dataflow. The rates matrix $\Gamma_r$ gives the rate of data production and consumption normalized to the interval $[0, 1]$, where $0$ indicates the absence of any dataflow, and $1$ indicates dataflow at the rate of one unit per cycle.

Building on the decomposed topology matrix formulation, a unified linear algebra framework is developed to formally characterize the design space explored by fpgaConvNet. Based on this unified framework, three combinatorial optimization problems are formulated subject to resource constraints on the targeted FPGA platform. These problems correspond to throughput optimization, latency optimization, and latency-constrained throughput optimization. Simulated annealing is applied to derive optimized designs based on the formulated design optimization problems.

The fpgaConvNet framework is demonstrated to generate implementations that provide higher performance density compared to state-of-the-art FPGA designs, and more favorable performance/power trade-offs compared to highly optimized designs on embedded GPUs. For further details, we refer the reader to~\cite{veni2019x1}.

\subsection{HOPES}

The Hope Of Parallel Embedded Software (HOPES) framework provides a design environment for embedded multiprocessor systems on chip (MPSoCs) based on integrated use of FSM, SDF, and SADF modeling techniques~\cite{ha2017x3}.
A particularly interesting feature in HOPES is it capability for robust scheduling of dynamically-varying application use cases~\cite{jung2014x1}. A {\em use case} in HOPES is defined as a set of distinct applications, where each application is represented as an SDF graph, and timing information is incorporated along with the SDF specifications. At any given time during execution of a system, only one use case is active; as execution evolves, the system can transition from executing one use case to another. 

In HOPES, SDF applications are compiled to a task model called common intermediate code (CIC)~\cite{kwon2008x1}, which is a graphical representation where, as in dataflow, vertices correspond to computational tasks and edges correspond to communication channels between tasks. CIC tasks are annotated with information to facilitate more efficient execution on embedded and parallel hardware. For example, data parallel tasks are annotated with their maximum degree of parallelism. Tasks in CIC correspond to threads that are managed by an operating system or by a runtime system that is synthesized by the HOPES framework. Tasks are either time driven or data driven. Time driven tasks are executed at specific times, whereas data driven tasks operate as dataflow actors, which execute based on the availability of sufficient input data.

The CIC model also has a concept of control tasks; the internal behavior of such a task is governed by an FSM. Each control task has an associated set of non-control tasks whose state it can manipulate. The FSMs associated with different control tasks in general execute concurrently rather than through the use of a monolithic FSM for the whole system. Examples of ways in which a control task controls another task are terminating, suspending, or resuming the task or changing parameters of the task. Timing requirements are also managed within control tasks.

CIC tasks in general have two levels of hierarchy, where a given task may include a nested SDF graph. The front-end to the framework may contain additional levels of hierarchy, but during synthesis, such hierarchy is flattened to the two-level form of hierarchy imposed by the CIC model.

An important function of control tasks in HOPES is that they are used to model inter-application dynamics. To model intra-application dynamics, an extension of the CIC model is used that is based on a restricted form of SADF (see \mxsecref{FSMintegration} and \mxsecref{videoExamples}) called FSM-based SADF~\cite{stui2011x1}. In particular, individual modes of a dynamic task are specified as an SDF graph, and an FSM controls the transitions that determine which mode is active at a given time during execution.

HOPES features a hybrid compile-time/run-time mapping technique to provide efficient and robust execution of dynamic application behavior and to address dynamic changes in resource availability, such as those that arise due to processor failures. In this context, an application can be viewed as a composite CIC task, where, as described above, the internal functionality is specified as an SDF graph. The mapping technique is designed to optimize energy consumption subject to given throughput constraints on the tasks, and subject to constraints imposed by the set of available processors in the system. 

For each mode of each application in the overall system specification, static schedules for the associated SDF graph are constructed at compile-time for different numbers of available processor cores. More specifically, a set of schedules is derived that provides a Pareto front in the design evaluation space of energy consumption and number of allocated processors. A genetic algorithm approach developed by Marwedel et al.~is used for this purpose~\cite{marw2011x1}. 

A runtime manager executing on a control processor monitors the system state and performs dynamic modification of the processor allocation and task schedules in response to certain types of changes in the state.  Dynamic changes to system state that can trigger run-time modification to the allocation and schedule include changes to the use case (set of active applications) being executed and reductions in the set of available processing resources due to processor failures. The energy overhead for run-time management is considered in the framework, as well as the overhead incurred by remapping tasks to different processors when processor allocations are changed at run-time.

The HOPES framework along with its advanced methods for managing system dynamics are demonstrated concretely through a smartphone system in~\cite{jung2014x1}. The demonstration system is composed of several distinct application tasks, including a H.264, MP3 and G.723 decoders and x264 and G.723 encoders.

\section{Simulation}
\label{sec:simulation}

Simulation is very important in system-on-chip design. Simulation is not limited to functional verification, as with logic design. SoC designers use simulation to measure the performance and power consumption of their SoC designs. This is due in part to the fact that much of the functionality is implemented in software, which must be measured relative to the processors on which it runs. It is also due to the fact that the complex input patterns inherent in many SoC applications do not lend themselves to closed-form analysis. 

SystemC is a simulation language that is widely used to model systems-on-chips~\cite{blac2010x1}. SystemC leverages the C++ programming language to build a simulation environment. SystemC classes allow designers to describe a digital system using a combination of structural and functional techniques. SystemC supports simulation at several levels of abstraction. Register-transfer level simulations, for example, can be performed with the appropriate SystemC model. SystemC is most often used for more abstract models. A common type of model built in SystemC is a transaction-level model. This style of modeling describes the SoC as a network of communicating machines, with explicit connections between the models and functional descriptions for each model. The transaction-level model describes how data is moved between the models. 

Hardware/software co-simulators are multi-mode simulators that simultaneously simulate different parts of the system at different levels of detail. For example, some modules may be simulated in register-transfer mode while software running on a CPU is simulated functionally. Co-simulation is particularly useful for debugging the hardware/software interface, such as debugging driver software.  
 
Functional validation, performance analysis, and power analysis of systems-on-chips require simulating large numbers of vectors. Video and other SoC applications allow complex input sequences. Even relatively compact tests can take up tens of millions of bytes. These long input sequences are necessary to run the SoC through a reasonable number of the states implemented in the system. The large amounts of memory that can be integrated into today’s systems, whether they be on-chip or off-chip, allow the creation of SoCs with huge numbers of states that require long simulation runs.  

Simulators for software running on processors have been developed over the past several decades. The Synopsys Virtualizer, for example, provides a transaction-level modeling interface for software prototyping. Both computer architects and SoC designers need fast simulators to run the large benchmarks required to evaluate architectures. As a result, a number of simulation techniques covering a broad range of accuracy and performance have been developed. 

A simple method of analyzing a program’s execution behavior is to sample the program counter (PC) during program execution. The Unix prof command is an example of a PC-sampling analysis tool. PC sampling is subject to the same limitations on sampling rate as any other sampling process, but sampling rate is usually not a major concern in this case. A more serious limitation is that PC sampling gives us relative performance but not absolute performance. A sampled trace of the program counter tells us where the program spent its time during execution, which gives us valuable information about the relative execution time of program modules that can be used to optimize the program. But it does not give us the execution time on a particular platform---especially if the target platform is different than the platform on which the trace is taken---and so we must use other methods to determine the real-time performance of programs. 
Some simulators concentrate on the behavior of the cache, given the major role of the cache in determining overall system performance. 

The {\em dinero} simulator~\cite{edle2025x1} is a well-known example of a cache simulator. These simulators generally work from a trace generated from the execution of a program. The program to be analyzed is augmented with additional code that records the execution behavior of the program. The dinero simulator then reconstructs the cache behavior from the program trace. The architect can view the cache in various states or calculate cache statistics. 
Some simulation systems model the behavior of the processor itself. A functional CPU simulator models instruction execution and maintains the state of the programming model, that is, the set of registers visible to the programmer. The functional simulator does not, however, model the performance or energy consumption of the program’s execution. 

A cycle-accurate simulator of a CPU is designed to accurately predict the number of clock cycles required to execute every instruction, taking into account pipeline and memory system effects. The CPU model must therefore represent the internal structure of the CPU accurately enough to show how resources in the processor are used. The SimpleScalar simulation tool~\cite{burg1997x1} is a well-known toolkit for building cycle-accurate simulators. SimpleScalar allows a variety of processor models to be built by a combination of parameterization of existing models and linking new simulation modules into the framework. 

Power simulators are related to cycle-accurate simulators. Accurate power estimation requires models of the CPU microarchitecture at least as detailed as those used for performance evaluation. A power simulator must model all the important wires in the architecture since capacitance is a major source of power consumption. Wattch~\cite{broo2000x1} and SimplePower~\cite{vija2000x1} are the two best-known CPU power simulators. 

\section{Hardware/software Co-Synthesis}

Hardware/software co-synthesis tools allow system designers to explore architectural trade-offs. These tools take a description of a desired behavior that is relatively undifferentiated between hardware and software. They produce a heterogeneous hardware architecture and the architecture for the software to run on that platform. The software architecture includes the allocation of software tasks to the processing elements of the platform and the scheduling of computation and communication. 

The functional description of an application may take several forms. The most basic is a task graph, as shown in \mxfigref{taskGraph}. The graph describes data dependencies between a set of processes. Each component of the graph (that is, each set of connected nodes) forms a task. Each task runs periodically and every task can run at a different rate. The task graph model generally does not concern itself with the details of operations within a process.  The process is characterized by its execution time. Several variations of task graphs that include control information have been developed. In these models, the output of a process may enable one of several different processes.

\mxnewfig
\includegraphics[width=0.4\textwidth]
{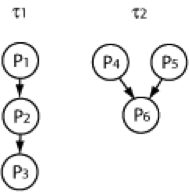}
\centering
 \caption{A task graph.}
 \label{fig:taskGraph}
\mxendfig

Task graph models are closely related to the dataflow graph models introduced in \mxsecref{dataflow}. The difference often lies in how the models are used. A key difference is that in dataflow models, emphasis is placed on precisely characterizing and analyzing how data is produced and consumed by computational components, while in task graph models, emphasis is placed on efficiently abstracting the execution time or resource utilization of the components or on analyzing real-time properties.  In many cases, dataflow graph techniques can be applied to the analysis or optimization of task graphs and vice versa. Thus, the terms ``task graph'' and ``dataflow graph'' are sometimes used interchangeably. 

An alternative representation for behavior is a programming language. Several different co-design languages have been developed and languages like SystemC have been used for co-synthesis as well. These languages may make use of constructs to describe parallelism that were originally developed for parallel programming languages. Such constructs are often used to capture operator-level concurrency. The subroutine structure of the program can be used to describe task-level parallelism. 

The most basic form of hardware/software co-synthesis is hardware/software partitioning. As shown in \mxfigref{template}, this method maps the design into an architectural template. The basic system architecture is bus-based, with a CPU and one or more custom hardware processing elements attached to the bus. The type of CPU is determined in advance, which allows the tool to accurately estimate software performance. The tool must decide what functions go into the custom processing elements; it must also schedule all the operations, whether implemented in hardware or software. This approach is known as hardware/software partitioning because the bus divides the architecture into two partitions and partitioning algorithms can be used to explore the design space.  

\mxnewfig
\includegraphics[width=0.4\textwidth]
{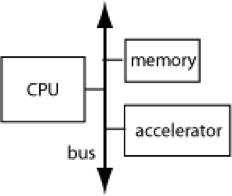}
\centering
 \caption{A template for hardware/software partitioning.}
 \label{fig:template}
\mxendfig

Two important approaches to searching the design space during partitioning were introduced by early tools. The Vulcan system~\cite{gupt1993x1} starts with all processes in custom processing elements and iteratively moves selected processes to the CPU to reduce the system cost. The COSYMA system~\cite{erns1993x1} starts with all operations running on the CPU and moves selected operations from loop nests into the custom processing element to increase performance. 
Hardware/software partitioning is ideally suited to platform FPGAs, which implement the bus-partitioned structure and use FPGA fabrics for the custom processing elements. However, the cost metric is somewhat different than in custom designs. Because the FPGA fabric is of a fixed size, using more or less of the fabric may not be important, so long as the design fits into the amount of logic available. 

Other co-synthesis algorithms have been developed that do not rely on an architectural template. Kalavade and Lee~\cite{kala1993x1} alternately optimize for performance and cost to generate a heterogeneous architecture. Wolf~\cite{wolf1997x1} alternated cost reduction and load balancing while maintaining a performance-feasible design. Dick and Jha~\cite{dick1998x1} used genetic algorithms to search the design space. 

Scheduling is an important task during co-synthesis. A complete system schedule must ultimately be constructed; an important aspect of scheduling is the scheduling of multiple processes on a single CPU. The study of real-time scheduling for uniprocessors was initiated by Liu and Layland~\cite{liu1973x1}, who developed rate-monotonic scheduling (RMS) and earliest-deadline-first (EDF) scheduling. RMS and EDF are priority-based schedulers, which use priorities to determine which process to run next. Many co-synthesis systems use custom, state-based schedulers that determine the process to execute based upon the state of the system. 

Design estimation is an important aspect of co-synthesis. While some software characteristics may be determined by simulation, hardware characteristics are often estimated using high-level synthesis. Henkel and Ernst~\cite{henk1995x1} used forms of high-level synthesis algorithms to quickly synthesize a hardware accelerator unit and estimate its performance and size.  Fornaciari et al.~\cite{forn1998x1} used high-level synthesis and libraries to estimate power consumption. 

Software properties may be estimated in a variety of ways, depending on the level of abstraction. For instance, Li and Wolf~\cite{li1999x1} built a process-level model of multiple processes interacting in the cache to provide an estimate of the performance penalty due to caches in a multi-tasking system. Tiwari et al.~\cite{tiwa1994x1} used measurements to build models of the power consumption of instructions executing on processors. 

\section{Machine Learning and System-Level Design}
\label{sec:learning}

Machine learning methods have seen increasing use in CAD, including system-level design.  
Several authors have studied the use of large language models (LLMs) for system-level design. Some LLMs have been trained on software and HDL code bases. LLM approaches must take into consideration the size and complexity of the HDL model to be generated. While small models may be generated directly from a prompt, generating a large HDL model may necessitate an iterative or hierarchical approach. 

Blocklove et al.~\cite{bloc2024x1} evaluated ChatGPT-4 on a set of benchmarks such as a shift register and traffic light controller. They found that the LLM generally performed well although its implementation of a dice roller did not generate pseudo-random results. Thakur et al.~\cite{thak2024x1} used an iterative approach to generate HDL from natural language prompts. They use results from compilation and simulation to generate additional prompt information to improve the design. They evaluated the approach using tutorial text from HDLbits. Batten et al.~\cite{batt2024x1} evaluated several LLMs, some used to generate Verilog and others used to generate Python-based hardware description languages. They found that the LLMs on Verilog generation was superior to that of Python generation. Several groups studied hierarchical prompting methods. Nakkab et al.~\cite{nakk2024x1} used hierarchical prompting: they extract a module hierarchy from the user’s problem description; they generate component modules one at a time and use simulation to improve the module; they then use the LLM to generate the top-level module from component modules and evaluate using simulation. VHDL-Xform~\cite{vija2024x1} uses a chain-of-descriptions approach to generate VHDL over a series of steps. A problem statement is used to generate a plan for creating the HDL for the desired function. The plan is streamlined to eliminate boilerplate text and ensure consistency across steps. The plan is then combined with the original prompt to cause the LLM to generate a design. Both single-step and multi-step approaches were considered. They found that multi-step methods gave improved results. A problem statement for code generation, VHDL code, and problem statement are given to an LLM to generate HDL.  

The generation of HDL designs introduces challenges beyond those seen for the generation of software by LLMs. An HDL design must be functionally correct, as for software, but must also meet non-functional requirements like clock speed and power consumption. Bai et al.~\cite{bai2024x1} explored methods to rank HDL designs and the effect of pragmas. They combined a pointwise prediction model for initial pruning of a design space.  They then used a pairwise comparison for more precise performance analysis. Qin et al.~\cite{qin2024x1} combine use of source code and control dataflow graph (CDFG) models of the HDL. Their large language model is trained to predict the quality of C/C++ HLS models and associated pragmas. The CDFG is used to augment the source code sequence. ReLS~\cite{lian2024x1} uses retrieval to enhance logic synthesis by combining new and legacy synthesis data. And-invert graphs are encoded along with synthesis recipes and their quality-of-result is predicted; the results are stored in a database. During logic synthesis, an and-invert graph for synthesis is used to retrieve a candidate recipe with high quality-of-result. The retrieved recipe is used to guide logic synthesis. 

Xu et al.~\cite{xu2024x1} explored the use of LLMs to repair C/C++ code during the process of conversion to high-level synthesis use. They used retrieval-augmented generation: they manually built a library of correction templates; they then used the LLM to search the library and apply appropriate corrections. They developed approaches to optimize bit widths and to add or tune pragmas.

\section{Summary}

System-level design is challenging because it is heterogeneous. The applications that we want to implement are heterogeneous in their computational models. The architectures on which we implement these applications are also heterogeneous combinations of custom hardware, processors, and memory. As a result, system-level tools help designers manage and understand complex, heterogeneous systems. Models of computation help designers cast their problem in a way that can be clearly understood by both humans and tools. Simulation helps designers gather important design characteristics. Hardware/software co-synthesis helps explore design spaces. As applications become more complex, expect to see tools continue to reach into the application space to aid with the transition from algorithm to architecture. 

\bibliographystyle{unsrt} 
\bibliography{refs}

\begin{thebibliography}{100}

\bibitem{frit2000x1}
J.~Fritts and W.~Wolf.
\newblock Evaluation of static and dynamic scheduling for media processors.
\newblock In {\em Proceedings of the MICRO-33 MP-DSP2 Workshop}, Monterey, California, 2000.

\bibitem{tall2000x1}
D.~Talla, L.~K. John, V.~Lapinskii, and B.~L. Evans.
\newblock Evaluating signal processing and multimedia applications on {SIMD}, {VLIW} and superscalar architectures.
\newblock In {\em Proceedings of the International Conference on Computer Design}, pages 163--172, Austin, Texas, 2000.

\bibitem{rang2021x1}
P.~Ranganathan et~al.
\newblock Warehouse-scale video acceleration: co-design and deployment in the wild.
\newblock In {\em Proceedings of the ACM International Conference on Architectural Support for Programming Languages and Operating Systems}, pages 600--615, Lausanne, Switzerland, 2021.

\bibitem{qual2014x1}
Qualcomm, Inc.
\newblock {\em The {Qualcomm} {Snapdragon} 800}, 2014.
\newblock \url{http://www.qualcomm.com/snapdragon/processors/800}, accessed 21 April 2014.

\bibitem{free2012x1}
Freescale Semiconductor.
\newblock {\em {MPC5676R} Microcontroller Data Sheet}, September 2012.
\newblock document number MPC5676R, rev. 3.

\bibitem{nvid2009x1}
NVIDIA Corporation.
\newblock {\em Whitepaper: {NVIDIA's} Next Generation {CUDA} Compute Architecture: Fermi}, 2009.

\bibitem{wolf2004x1}
W.~Wolf.
\newblock {\em {FPGA}-Based System Design}.
\newblock Prentice Hall, 2004.

\bibitem{xili2014x1}
Xilinx, Inc.
\newblock {\em Zynq-7000 All Programmable {SoC} Technical Reference Manual}, September 2014.

\bibitem{alte2014x1}
Altera, Inc.
\newblock {\em Arria 10 Device Overview}, 2014.

\bibitem{cypr2013x1}
Cypress Semiconductor.
\newblock {\em {PSoC} {5LP} Architecture Technical Reference Manual}, 2013.
\newblock Document no. 001-78426.

\bibitem{joup2017x1}
N.~P. Jouppi et~al.
\newblock In-datacenter performance analysis of a tensor processing unit.
\newblock In {\em International Symposium on Computer Architecture}, pages 1--12, Toronto, Canada, 2017.

\bibitem{gajs1983x1}
D.~D. Gajski and R.~H. Kuhn.
\newblock Guest editors' introduction: New {VLSI} tools.
\newblock {\em Computer}, 16(12):11--14, 1983.

\bibitem{gers2009x1}
A.~Gerstlauer, C.~Haubelt, A.~D. Pimentel, T.~P. Stefanov, D.~D. Gajski, and J.~Teich.
\newblock Electronic system-level synthesis methodologies.
\newblock {\em IEEE Transactions on Computer-Aided Design of Integrated Circuits and Systems}, 28(10):1517--1530, 2009.

\bibitem{lee1998x3}
E.~A. Lee and A.~Sangiovanni-Vincentelli.
\newblock A framework for comparing models of computation.
\newblock {\em IEEE Transactions on Computer-Aided Design of Integrated Circuits and Systems}, 17(12):1217--1229, December 1998.

\bibitem{basu1997x1}
A.~Basu, M.~Hayden, G.~Morrisett, and T.~{von Eicken}.
\newblock A language-based approach to protocol construction.
\newblock In {\em ACM SIGPLAN Workshop on Domain-Specific Languages}, Paris, France, 1997.

\bibitem{kons1994x1}
K.~Konstantinides and J.~R. Rasure.
\newblock The {Khoros} software-development environment for image-processing and signal-processing.
\newblock {\em IEEE Transactions on Image Processing}, 3(3):243--252, May 1994.

\bibitem{lauw1995x1}
R.~Lauwereins, M.~Engels, M.~Ade, and J.~A. Peperstraete.
\newblock Grape-{II}: A system-level prototyping environment for {DSP} applications.
\newblock {\em Computer}, 28(2):35--43, February 1995.

\bibitem{lee1989x2}
E.~A. Lee, W.~H. Ho, E.~Goei, J.~Bier, and S.~S. Bhattacharyya.
\newblock Gabriel: A design environment for {DSP}.
\newblock {\em IEEE Transactions on Acoustics, Speech, and Signal Processing}, 37(11):1751--1762, November 1989.

\bibitem{conw2004x1}
C.~L. Conway and S.~A. Edwards.
\newblock {NDL}: A domain-specific language for device drivers.
\newblock In {\em Proceedings of the Workshop on Languages Compilers and Tools for Embedded Systems}, Washington, D.C., 2004.

\bibitem{mani1998x2}
V.~Manikonda, P.S. Krishnaprasad, and J.~Hendler.
\newblock Languages, behaviors, hybrid architectures and motion control.
\newblock Technical Report 98-3, University of Maryland Institute for Systems Research, 1998.

\bibitem{prou2001x1}
K.~Proudfoot, W.~R. Mark, S.~Tzvetkov, and P.~Hanrahan.
\newblock A real-time procedural shading system for programmable graphics hardware.
\newblock In {\em Proceedings of SIGGRAPH}, 2001.

\bibitem{thib1999x1}
S.~A. Thibault, R.~Marlet, and C.~Consel.
\newblock Domain-specific languages: From design to implementation application to video device drivers generation.
\newblock {\em IEEE Transactions on Software Engineering}, 25(3):363--377, May/June 1999.

\bibitem{bhat2019x1}
S.~S. Bhattacharyya, E.~Deprettere, R.~Leupers, and J.~Takala, editors.
\newblock {\em Handbook of Signal Processing Systems}.
\newblock Springer, third edition, 2019.

\bibitem{eker2012x1}
J.~Eker and J.~W. Janneck.
\newblock Dataflow programming in {CAL} --- balancing expressiveness, analyzability, and implementability.
\newblock In {\em Proceedings of the IEEE Asilomar Conference on Signals, Systems, and Computers}, pages 1120--1124, Pacific Grove, California, 2012.

\bibitem{pelc2013x1}
M.~Pelcat, S.~Aridhi, J.~Piat, and J.-F. Nezan.
\newblock {\em Physical Layer Multi-Core Prototyping}.
\newblock Springer, 2013.

\bibitem{bhat2008x4}
S.~S. Bhattacharyya, G.~Brebner, J.~Eker, J.~W. Janneck, M.~Mattavelli, C.~{von Platen}, and M.~Raulet.
\newblock {OpenDF} --- a dataflow toolset for reconfigurable hardware and multicore systems.
\newblock In {\em Proceedings of the Swedish Workshop on Multi-Core Computing}, pages 43--49, Ronneby, Sweden, 2008.

\bibitem{edwa2000x1}
S.~A. Edwards.
\newblock {\em Languages for Digital Embedded Systems}.
\newblock Kluwer Academic Publishers, 2000.

\bibitem{jant2003x1}
A.~Jantsch.
\newblock {\em Modeling Embedded Systems and {SoC's}: Concurrency and Time in Models of Computation}.
\newblock Morgan Kaufmann Publishers Inc., 2003.

\bibitem{karp1966x1}
R.~M. Karp and R.~E. Miller.
\newblock Properties of a model for parallel computations: Determinacy, termination, queuing.
\newblock {\em SIAM Journal of Applied Math}, 14(6), November 1966.

\bibitem{kahn1974x1}
G.~Kahn.
\newblock The semantics of a simple language for parallel programming.
\newblock In {\em Proceedings of the IFIP Congress}, Stockholm, Sweden, 1974.

\bibitem{lee1995x1}
E.~A. Lee and T.~M. Parks.
\newblock Dataflow process networks.
\newblock {\em Proceedings of the IEEE}, 83(5):773--799, 1995.

\bibitem{ambl1992x1}
A.~L. Ambler, M.~M. Burnett, and B.~A. Zimmerman.
\newblock Operational versus definitional: A perspective on programming paradigms.
\newblock {\em Computer}, 25(9):28--43, September 1992.

\bibitem{ha1991x1}
S.~Ha and E.~A. Lee.
\newblock Compile-time scheduling and assignment of data-flow program graphs with data-dependent iteration.
\newblock {\em IEEE Transactions on Computers}, 40(11):1225--1238, November 1991.

\bibitem{eker2003x4}
J.~Eker et~al.
\newblock Taming heterogeneity --- the {Ptolemy} approach.
\newblock {\em Proceedings of the IEEE}, 91(1):127--144, 2003.

\bibitem{ko2005x1}
D.~Ko and S.~S. Bhattacharyya.
\newblock Modeling of block-based {DSP} systems.
\newblock {\em Journal of VLSI Signal Processing Systems for Signal, Image, and Video Technology}, 40(3):289--299, July 2005.

\bibitem{lee1987x1}
E.~A. Lee and D.~G. Messerschmitt.
\newblock Synchronous dataflow.
\newblock {\em Proceedings of the IEEE}, 75(9):1235--1245, September 1987.

\bibitem{lubl2008x2}
R.~Lublinerman and S.~Tripakis.
\newblock Translating data flow to synchronous block diagrams.
\newblock In {\em Proceedings of the IEEE Workshop on Embedded Systems for Real-Time Multimedia}, Atlanta, Georgia, 2008.

\bibitem{murt2002x1}
P.~K. Murthy and E.~A. Lee.
\newblock Multidimensional synchronous dataflow.
\newblock {\em IEEE Transactions on Signal Processing}, 50(8):2064--2079, August 2002.

\bibitem{vaid1993x1}
P.~P. Vaidyanathan.
\newblock {\em Multirate Systems and Filter Banks}.
\newblock Prentice Hall, 1993.

\bibitem{stic2002x2}
D.~Stichling and B.~Kleinjohann.
\newblock {CV-SDF} --- a synchronous data flow model for real-time computer vision applications.
\newblock In {\em Proceedings of the International Workshop on Systems, Signals and Image Processing}, Manchester, United Kingdom, 2002.

\bibitem{kein2006x1}
J.~Keinert, C.~Haubelt, and J.~Teich.
\newblock Modeling and analysis of windowed synchronous algorithms.
\newblock In {\em Proceedings of the International Conference on Acoustics, Speech, and Signal Processing}, Toulouse, France, 2006.

\bibitem{buck1993x1}
J.~T. Buck and E.~A. Lee.
\newblock Scheduling dynamic dataflow graphs with bounded memory using the token flow model.
\newblock In {\em Proceedings of the International Conference on Acoustics, Speech, and Signal Processing}, Minneapolis, Minnesota, April 1993.

\bibitem{buck1994x2}
J.~T. Buck.
\newblock Static scheduling and code generation from dynamic dataflow graphs with integer-valued control streams.
\newblock In {\em Proceedings of the IEEE Asilomar Conference on Signals, Systems, and Computers}, pages 508--513, Pacific Grove, California, October 1994.

\bibitem{buck2000x1}
J.~Buck and R.~Vaidyanathan.
\newblock Heterogeneous modeling and simulation of embedded systems in {El Greco}.
\newblock In {\em Proceedings of the International Workshop on Hardware/Software Codesign}, San Diego, California, 2000.

\bibitem{gira1999x1}
A.~Girault, B.~Lee, and E.~A. Lee.
\newblock Hierarchical finite state machines with multiple concurrency models.
\newblock {\em IEEE Transactions on Computer-Aided Design of Integrated Circuits and Systems}, 18(6):742--760, June 1999.

\bibitem{thie1999x1}
L.~Thiele, K.~Strehl, D.~Ziegenbein, R.~Ernst, and J.~Teich.
\newblock {FunState} --- an internal representation for codesign.
\newblock In {\em Proceedings of the International Conference on Computer-Aided Design}, San Jose, California, 1999.

\bibitem{coss2000x1}
N.~Cossement, R.~Lauwereins, and F.~Catthoor.
\newblock {DF*}: An extension of synchronous dataflow with data dependency and non-determinism.
\newblock In {\em Proceedings of the Forum on Specification and Design Languages}, Tuebingen, Germany, 2000.

\bibitem{pank1994x1}
M.~Pankert, O.~Mauss, S.~Ritz, and H.~Meyr.
\newblock Dynamic data flow and control flow in high level {DSP} code synthesis.
\newblock In {\em Proceedings of the International Conference on Acoustics, Speech, and Signal Processing}, Adelaide, Australia, 1994.

\bibitem{plis2008x5}
W.~Plishker, N.~Sane, M.~Kiemb, K.~Anand, and S.~S. Bhattacharyya.
\newblock Functional {DIF} for rapid prototyping.
\newblock In {\em Proceedings of the International Symposium on Rapid System Prototyping}, pages 17--23, Monterey, California, 2008.

\bibitem{thee2006x1}
B.~D. Theelen, M.~C.~W. Geilen, T.~Basten, J.~P.~M. Voeten, S.~V. Gheorghita, and S.~Stuijk.
\newblock A scenario-aware data flow model for combined long-run average and worst-case performance analysis.
\newblock In {\em Proceedings of the International Conference on Formal Methods and Models for Codesign}, Napa, California, 2006.

\bibitem{jann2010x1}
J.~W. Janneck, M.~Mattavelli, M.~Raulet, and M.~Wipliez.
\newblock Reconfigurable video coding: a stream programming approach to the specification of new video coding standards.
\newblock In {\em Proceedings of the ACM SIGMM conference on Multimedia systems}, pages 223--234, Florence, Italy, 2010.

\bibitem{eker2003x2}
J.~Eker and J.~W. Janneck.
\newblock {CAL} language report, language version 1.0 --- document edition 1.
\newblock Technical Report UCB/ERL M03/48, Electronics Research Laboratory, University of California at Berkeley, December 2003.

\bibitem{yviq2013x2}
H.~Yviquel et~al.
\newblock Orcc: multimedia development made easy.
\newblock In {\em Proceedings of the ACM International Conference on Multimedia}, pages 863--866, Barcelona, Spain, 2013.

\bibitem{bout2008x1}
J.~Boutellier, V.~Sadhanala, C.~Lucarz, P.~Brisk, and M.~Mattavelli.
\newblock Scheduling of dataflow models within the reconfigurable video coding framework.
\newblock In {\em Proceedings of the IEEE Workshop on Signal Processing Systems}, Washington, D.C., 2008.

\bibitem{gu2011x2}
R.~Gu et~al.
\newblock Exploiting statically schedulable regions in dataflow programs.
\newblock {\em Journal of Signal Processing Systems}, 63(1):129--142, 2011.

\bibitem{stef2004x1}
T.~Stefanov, C.~Zissulescu, A.~Turjan, B.~Kienhuis, and E.~Deprettere.
\newblock System design using {Kahn} process networks: the {Compaan}/{Laura} approach.
\newblock In {\em Proceedings of the Design, Automation and Test in Europe Conference and Exhibition}, Paris, France, 2004.

\bibitem{bils1996x1}
G.~Bilsen, M.~Engels, R.~Lauwereins, and J.~A. Peperstraete.
\newblock Cyclo-static dataflow.
\newblock {\em IEEE Transactions on Signal Processing}, 44(2):397--408, February 1996.

\bibitem{depr2006x1}
E.~F. Deprettere, T.~Stefanov, S.~S. Bhattacharyya, and M.~Sen.
\newblock Affine nested loop programs and their binary cyclo-static dataflow counterparts.
\newblock In {\em Proceedings of the International Conference on Application Specific Systems, Architectures, and Processors}, pages 186--190, Steamboat Springs, Colorado, 2006.

\bibitem{kien2000x1}
B.~Kienhuis, E.~Rijpkema, and E.~Deprettere.
\newblock Compaan: deriving process networks from {Matlab} for embedded signal processing architectures.
\newblock In {\em Proceedings of the International Workshop on Hardware/Software Codesign}, San Diego, California, 2000.

\bibitem{turj2004x1}
A.~Turjan, B.~Kienhuis, and E.~Deprettere.
\newblock An integer linear programming approach to classify the communication in process networks.
\newblock In {\em Proceedings of the International Workshop on Software and Compilers for Embedded Systems}, pages 62--76, Amsterdam, The Netherlands, September 2004.

\bibitem{niko2008x2}
H.~Nikolov, M.~Thompson, T.~Stefanov, A.~Pimentel, S.~Polstra, R.~Bose, C.~Zissulescu, and E.~Deprettere.
\newblock Daedalus: toward composable multimedia {MP-SoC} design.
\newblock In {\em Proceedings of the Design Automation Conference}, pages 574--579, Anaheim, California, 2008.

\bibitem{pelc2009x2}
M.~Pelcat, J.~Piat, M.~Wipliez, S.~Aridhi, and J.-F. Nezan.
\newblock An open framework for rapid prototyping of signal processing applications.
\newblock {\em EURASIP Journal on Embedded Systems}, 2009, January 2009.
\newblock Article No. 11.

\bibitem{holz2004x1}
S.~Holzner.
\newblock {\em Eclipse}.
\newblock O'Reilly \& Associates, Inc., 2004.

\bibitem{pelc2014x2}
M.~Pelcat et~al.
\newblock Dataflow-based rapid prototyping for multicore {DSP} systems.
\newblock Technical Report PREESM/2014-05TR01, Institut National des Sciences Appliqu\'{e}es de Rennes, 2014.

\bibitem{buck1994x1}
J.~T. Buck, S.~Ha, E.~A. Lee, and D.~G. Messerschmitt.
\newblock Ptolemy: A framework for simulating and prototyping heterogeneous systems.
\newblock {\em International Journal of Computer Simulation}, 4:155--182, April 1994.

\bibitem{kala1993x1}
A.~Kalavade and E.~A. Lee.
\newblock A hardware/software codesign methodology for {DSP} applications.
\newblock {\em IEEE Design \& Test of Computers}, 10(3):16--28, September 1993.

\bibitem{neue2004x2}
S.~Neuendorffer and E.~Lee.
\newblock Hierarchical reconfiguration of dataflow models.
\newblock In {\em Proceedings of the International Conference on Formal Methods and Models for Codesign}, San Diego, California, 2004.

\bibitem{zhou2007x1}
G.~Zhou, M.~Leung, and E.~A. Lee.
\newblock A code generation framework for actor-oriented models with partial evaluation.
\newblock Technical Report UCB/EECS-2007-29, Department of Electrical Engineering and Computer Sciences, University of California at Berkeley, February 2007.

\bibitem{alfa2001x1}
L.~de~Alfaro and T.~Henzinger.
\newblock Interface automata.
\newblock In {\em Proceedings of the Joint European Software Engineering Conference and {ACM SIGSOFT} International Symposium on the Foundations of Software Engineering}, Vienna Austria, 2001.

\bibitem{lee2001x4}
E.~A. Lee and Y.~Xiong.
\newblock System-level types for component-based design.
\newblock In {\em Proceedings of the International Workshop on Embedded Software}, pages 148--165, Tahoe City, California, October 2001.

\bibitem{haub2007x1}
C.~Haubelt, J.~Falk, J.~Keinert, T.~Schlichter, M.~{Streub\"{u}hr}, A.~Deyhle, A.~Hadert, and J.~Teich.
\newblock A {SystemC}-based design methodology for digital signal processing systems.
\newblock {\em EURASIP Journal on Embedded Systems}, 2007:Article ID 47580, 22 pages, 2007.

\bibitem{veni2019x1}
S.~I. Venieris and C.-S. Bouganis.
\newblock {fpgaConvNet}: Mapping regular and irregular convolutional neural networks on {FPGAs}.
\newblock {\em IEEE Transactions on Neural Networks and Learning Systems}, 30(2):326--342, 2019.

\bibitem{ha2017x3}
S.~Ha and H.~Jung.
\newblock {HOPES}: Programming platform approach for embedded systems design.
\newblock In S.~Ha and J.~Teich, editors, {\em Handbook of hardware/software codesign}. Springer Dordrecht, 2017.

\bibitem{jung2014x1}
H.~Jung, C.~Lee, S.-H. Kang, S.~Kim, H.~Oh, and S.~Ha.
\newblock Dynamic behavior specification and dynamic mapping for real-time embedded systems: {HOPES} approach.
\newblock {\em ACM Transactions on Embedded Computing Systems}, 13(4s):1--26, 2014.

\bibitem{kwon2008x1}
S.~Kwon, Y.~Kim, {W.-C.} Jeun, S.~Ha, and Y.~Paek.
\newblock A retargetable parallel-programming framework for {MPSoC}.
\newblock {\em ACM Transactions on Design Automation of Electronic Systems}, 13(3), July 2008.

\bibitem{stui2011x1}
S.~Stuijk, M.~Geilen, B.~Theelen, and Twan Basten.
\newblock Scenario-aware dataflow: Modeling, analysis and implementation of dynamic applications.
\newblock In {\em Proceedings of the International Conference on Embedded Computer Systems: Architectures, Modeling, and Simulation}, pages 404--411, Samos, Greece, 2011.

\bibitem{marw2011x1}
P.~Marwedel, J.~Teich, G.~Kouveli, J.~Bacivarov, L.~Thiele, S.~Ha, C.~Lee, Q.~Xu, and L.~Huang.
\newblock Mapping of applications to {MPSoCs}.
\newblock In {\em Proceedings of the International Conference on Hardware/Software Codesign and System Synthesis}, pages 109--118, Taipei, Taiwan, 2011.

\bibitem{blac2010x1}
D.~C. Black, J.~Donovan, B.~Bunton, and A.~Keist.
\newblock {\em {SystemC}: From the Ground Up}.
\newblock Springer, second edition, 2010.

\bibitem{edle2025x1}
J.~Edler and M.~D. Hill.
\newblock Dinero {IV} trace-driven uniprocessor cache simulator.
\newblock University of Wisconsin, \url{https://pages.cs.wisc.edu/~markhill/DineroIV/}, Visited on June 1, 2025.

\bibitem{burg1997x1}
D.~C. Burger and T.~M. Austin.
\newblock The simplescalar tool set, version 2.0.
\newblock Technical Report 1342, Department of Computer Sciences, University of Wisconsin at Madison, June 1997.

\bibitem{broo2000x1}
D.~Brooks, V.~Tiwari, and M.~Martonosi.
\newblock Wattch: a framework for architectural-level power analysis and optimizations.
\newblock In {\em International Symposium on Computer Architecture}, pages 83--94, Vancouver, Canada, 2000.

\bibitem{vija2000x1}
N.~Vijaykrishnan, M.~Kandemir, M.~J. Irwin, H.~S. Kim, and W.~Ye.
\newblock Energy-driven integrated hardware-software optimizations using {SimplePower}.
\newblock In {\em International Symposium on Computer Architecture}, pages 95--106, Vancouver, Canada, 2000.

\bibitem{gupt1993x1}
R.~K. Gupta and G.~{De Micheli}.
\newblock Hardware-software cosynthesis for digital systems.
\newblock {\em IEEE Design \& Test of Computers}, 10(3):29--41, 1993.

\bibitem{erns1993x1}
R.~Ernst, J.~Henkel, and T.~Benner.
\newblock Hardware-software cosynthesis for microcontrollers.
\newblock {\em IEEE Design \& Test of Computers}, 10(4):64--75, December 1993.

\bibitem{wolf1997x1}
W.~H. Wolf.
\newblock An architectural co-synthesis algorithm for distributed, embedded computing systems.
\newblock {\em IEEE Transactions on Very Large Scale Integration (VLSI) Systems}, 5(2):218--229, 1997.

\bibitem{dick1998x1}
R.~P. Dick and N.~K. Jha.
\newblock {MOGAC}: A multiobjective genetic algorithm for hardware-software cosynthesis of distributed embedded systems.
\newblock {\em IEEE Transactions on Computer-Aided Design of Integrated Circuits and Systems}, 17(10):920--935, October 1998.

\bibitem{liu1973x1}
C.~L. Liu and J.~W. Layland.
\newblock Scheduling algorithms for multiprogramming in a hard-real-time environment.
\newblock {\em Journal of the Association for Computing Machinery}, 20(1):46--61, 1973.

\bibitem{henk1995x1}
J.~Henkel and R.~Ernst.
\newblock A path-based technique for estimating hardware runtime in {HW/SW}-cosynthesis.
\newblock In {\em Proceedings of the International Symposium on System Synthesis}, pages 116--121, Cannes, France, 1995.

\bibitem{forn1998x1}
W.~Fornaciari, P.~Gubian, D.~Sciuto, and C.~Silvano.
\newblock Power estimation of embedded systems: A hardware/software codesign approach.
\newblock {\em IEEE Transactions on Very Large Scale Integration (VLSI) Systems}, 6(2):266--275, June 1998.

\bibitem{li1999x1}
Y.~Li and W.~Wolf.
\newblock Hardware/software co-synthesis with memory hierarchies.
\newblock {\em IEEE Transactions on Computer-Aided Design of Integrated Circuits and Systems}, 18(10):1405--1417, October 1999.

\bibitem{tiwa1994x1}
V.~Tiwari, S.~Malik, and A.~Wolfe.
\newblock Power analysis of embedded software: a first step towards software power minimization.
\newblock {\em IEEE Transactions on Very Large Scale Integration (VLSI) Systems}, December 1994.

\bibitem{bloc2024x1}
J.~Blocklove, S.~Garg, R.~Karri, and H.~Pearce.
\newblock Evaluating {LLMs} for hardware design and test, 2024.
\newblock arXiv:2405.02326v2 [cs.AR].

\bibitem{thak2024x1}
S.~Thakur, J.~Blocklove, H.~Pearce, B.~Tan, S.~Garg, and R.~Karri.
\newblock Autochip: Automating {HDL} generation using {LLM} feedback, 2024.
\newblock arXiv:2311.04887v2 [cs.PL].

\bibitem{batt2024x1}
C.~Batten, N.~Pinckney, M.~Liu, H.~Ren, and B.~Khailany.
\newblock {PyHDL-Eval}: An {LLM} evaluation framework for hardware design using {Python}-embedded {DSLs}.
\newblock In {\em Proceedings of the ACM/IEEE International Symposium on Machine Learning for CAD}, pages 1--17, Snowbird, Utah, 2024.

\bibitem{nakk2024x1}
A.~Nakkab, S.~Q. Zhang, R.~Karri, and S.~Garg.
\newblock Rome was not built in a single step: Hierarchical prompting for {LLM}-based chip design.
\newblock In {\em Proceedings of the ACM/IEEE International Symposium on Machine Learning for CAD}, pages 1--11, Snowbird, Utah, 2024.

\bibitem{vija2024x1}
P.~Vijayaraghavan et~al.
\newblock Chain-of-descriptions: Improving code {LLMs} for {VHDL} code generation and summarization.
\newblock In {\em Proceedings of the ACM/IEEE International Symposium on Machine Learning for CAD}, pages 1--10, Snowbird, Utah, 2024.

\bibitem{bai2024x1}
Y.~Bai et~al.
\newblock Learning to compare hardware designs for high-level synthesis.
\newblock In {\em Proceedings of the ACM/IEEE International Symposium on Machine Learning for CAD}, pages 1--7, Snowbird, Utah, 2024.

\bibitem{qin2024x1}
Z.~Qin et~al.
\newblock Cross-modality program representation learning for electronic design automation with high-level synthesis.
\newblock In {\em Proceedings of the ACM/IEEE International Symposium on Machine Learning for CAD}, pages 1--12, Snowbird, Utah, 2024.

\bibitem{lian2024x1}
R.~Liang, C.-T. Ho, A.~Agnesina, W.-H. Liu, and H.~Ren.
\newblock {ReLS}: Retrieval is efficient knowledge transfer for logic synthesis.
\newblock In {\em Proceedings of the ACM/IEEE International Symposium on Machine Learning for CAD}, pages 1--7, Snowbird, Utah, 2024.

\bibitem{xu2024x1}
K.~Xu et~al.
\newblock Automated {C/C++} program repair for high-level synthesis via large language models.
\newblock In {\em Proceedings of the ACM/IEEE International Symposium on Machine Learning for CAD}, pages 1--9, Snowbird, Utah, 2024.

\end{thebibliography}
\end{document}